**Beyond the biology: Evaluating the role of political, economic, and social factors associated with the incidence and mortality of SARS-CoV-2 during the first seven months of the COVID-19 pandemic**

Elizabeth A. Schafer [1], Samantha Eberl [2] and Lauro Velazquez-Salinas [3*]

Affiliations

1. Tunnell Government Services Inc., Bethesda, MD 20817, USA

2. Department of Psychological Science, Central Connecticut State University, New Britain, CT 06053, USA.

3. College of Veterinary Medicine, Kansas State University, Manhattan, KS 66506, USA.

* Corresponding author: Lauro Velazquez-Salinas. Lauro.velazquez@usda.gov   College of Veterinary Medicine, Kansas State University, Manhattan, KS 66506, USA

**Abstract**

Purpose: In this study we propose to identify significant economic, political, and social factors that helped contribute to the incidence and mortality of SARS-CoV-2 around the world during the first seven months of the COVID-19 pandemic.

Methods: Retrospective analyses were conducted using data acquired by the World Health Organization regarding the number of new cases and deaths from 76 countries from January through July 2020. Hierarchical cluster and principal component analyses were used to categorize different countries based on their ability to control the outbreak based on number of cases and deaths throughout these initial months of the pandemic. Furthermore, using non-parametrical analyses, different variables were correlated with multiple economic, political, and social indexes.

Results: Multiple significant correlations were found, including the control of corruption, government effectiveness, and the human life indicator, with indexes containing the higher correlations demonstrating the capability of individual countries to help drive the course of this pandemic. Geographical hotspots were identified that required further attention to aid in outbreak control.

Conclusion: Collectively, the findings in this study indicate that multiple factors beyond the biological should be considered in the implementation of measures to contain this pandemic.

**Introduction**

SARS-CoV-2 emerged in 2019 to become the seventh human coronavirus documented in the world (Ye et al., 2020). Rapidly after the first cases were observed in China, SARS-CoV-2 quickly spread around the world, the beginning of what would soon be known as the COVID-19 pandemic of 2020 (Wu et al., 2020). As the catastrophic impact of the pandemic progressed, the varied and unanticipated clinical features of SARS-CoV-2 lead to collapsed health systems and exposed deficiencies throughout the world (Armocida et al., 2020).

The initial approaches taken by different governments in order to mitigate the effects of the pandemic varied in general during the initial course of the pandemic; however, countries adopted the overall guideless proposed by the World Health Organization (WHO); (Alanezi et al., 2020). Additional policies included the implementation of lockdown (Alfano and Ercolano, 2020), the mandatory use of masks, a focus increasing the strength of the healthcare systems (https://eurohealthobservatory.who.int/publications/i/9999-12-31-eurohealth-health-systemsresponses-to-covid-19-26-2-) and early attempts at herd immunity (Alanezi et al., 2020).

Within the first seven months of the pandemic, interesting and dramatic shifts in terms of the incidence of SARS-CoV-2 were documented worldwide (Figures 1A and B). Initially (January of 2020), there was an increase of cases and deaths in the Western Pacific area, shifting just months later with (February to April), with Europe becoming the center of the pandemic. At this point, overall higher mortality rates were being registered around the world (Figure 1C). By the end of May, the number of cases in the Americas (2,698,714) were already slightly higher than Europe (2,119,383), the Americas becoming the new center of the pandemic. Interestingly, by the end of July, the differences in the number of cases between the Americas (9,152,173) and Europe (3,352,949) were extremely contrasting. Also, at this point of the pandemic, the South-East Asia region significantly increased the number of cases between June (272,512) and July (2,009,963). (https://covid19.who.int/).

Based on the above, the overall situation regarding the incidence and mortality of SARS-CoV-2 worldwide raises interesting questions about the existence of other factors to consider beyond the clinical and epidemiological characteristics of SARS-CoV-2 within this first wave of the outbreak (Wang et al., 2020). By taking a closer look at the influence of societal, political, and economic factors, one may be able to overall assess additional factors to the extent of the outbreak throughout different regions of the world. Indeed, this view is consistent with the result of previous studies (Mazzucchelli et al., 2020; Ojong, 2020; Quinn and Kumar, 2014; Thevenot, 2017)

Herein, we tested the hypothesis that political, economic, and social factors might have impaired the ability of different countries to control the incidence and mortality produced by SARS-CoV-2 during the first seven months of the pandemic. For this purpose, we establish multiple correlation analyses between the number of new cases and deaths from different countries and multiple social, political, and economic indexes. The results of this research illustrate the complexity of this pandemic event in terms of its control, showing that multiple factors beyond the biological characteristics of SARS-CoV-2 should also be considered in the elaboration of strategies to mitigate the effects of this disease around the world.

**Material and Methods**

Data Collection

All available data regarding the number of new cases and deaths in countries affected by COVID-19 were obtained from the database of the World Health Organization (https://covid19.who.int/) using available data from 1/01/20 until 7/31/2020. To increase the accuracy of our results, countries with 10, 000 or greater number of total cases (n=72) were chosen for purposes of this analysis. Four additional countries (<10,000 cases) were included as controls for our study based on their ability to control the initial phase of the outbreak (Finland, New Zealand, Iceland and Uruguay) (Cousins, 2020; Gudbjartsson et al., 2020; Moreno et al., 2020). For information about the countries and a detailed description of the different indexes included in this study, see supplementary file 1.

Spearman's Rank Correlation Coefficient (Spearman's $\rho$)

Countries were categorized based on their ability to control the outbreak with regard to mortality and number of infections and three different non-parametric correlation coefficients were identified using the Spearman's $\rho$ test (Bishara and Hittner, 2017). To establish the correlation of new cases with the number of new deaths in each country, the strength of the correlation between these two variables was determined. Next, in order to investigate the extent to which each country was able to control the outbreak based on case numbers and death, correlations were calculated between both the number of cases and time, and the number of deaths and time. Only correlations with p-values <0.0001 were considered significant.

Spearman's $\rho$ calculations were conducted on the software GraphPad Prism version 8.0.0 for Windows, GraphPad Software, San Diego, California USA, www.graphpad.com.

Hierarchical Cluster Analysis (HCA)

Countries were categorized using HCA (Velazquez-Salinas et al., 2016), classifying objects into clusters based on their similarities, using a matrix of N x M dimensions where N represents the number of countries and M the values of the three different correlation coefficients (cases/deaths, cases/time, and deaths/time). Data was robustly standardized, and the distance metrics between clusters was defined by the complete linkage method. The resultant inference was presented as a dendrogram, using a heat map to show the values of different correlation coefficients for each country.

Principal Component Analysis (PCA)

PCA was used to generate a coordinate system to numerically reflect the relationship among different countries based on the variables included in this study (cases/deaths, cases/time, and deaths/time). PCA classified different variables into main principal components, being the ones accounting for the highest variability associated with the first principal component (PC1) (Penny and Jolliffe, 1999). The number of significant components associated with PCA was validated by the Bartlett test (Bartlett, 1937). The association of different variables with specific principal components was determined by the Spearman's rank correlation coefficient.

Indexes

Finally, to determine potential political, economic, and social factors associated with the first principal component (PC1), Spearman's rank correlation coefficients were calculated between the coordinates of each country associated with the main PC1 and multiple indexes described below.

**Political indexes**: Control of corruption and government effectiveness were obtained from the Global Economy 2019 database (https://www.theglobaleconomy.com). The index of control of corruption captures the degree in which government power is used to favor private interests, while government effectiveness reflects the performance of the government to act independent of political pressures, evaluating the quality of policy formulation and implementation, and its commitment to such policies.

**Economical index**: Gross domestic product per capita (GDP per capita) information was obtained from the World Bank Data (https://data.worldbank.org/indicator/NY.GDP.MKTP.CD). This measurement reflects the economic output of a country divided by its total population during the year 2019.

**Social indexes**: The human development index (HDI;data 2016), the human life indicator (HLI;data 2016), and the life expectancy index (LE;data 2017) were obtained from the International Institute for Applied Systems Analysis (https://iiasa.ac.at/web/home/research/researchPrograms/WorldPopulation/Reaging/HLI.html).

HDI measures the level of human development at country level by considering three main components: health, education, and economic conditions. HLI accounts for wellbeing in terms of years of life, also considering inequality in longevity. LE is the arithmetic average of ages at death (Ghislandi et al., 2019).

Additionally, correlations between the PC1 and different risk factors associated with multiple causes of deaths were established (supplementary file 1; https://ourworldindata.org/causes-of-death),. Population data was obtained from the world population review database https://worldpopulationreview.com/

Statistical Analysis

HCA, PCA, and non-parametric correlations between PC1 and multiple indexes were conducted on the statistical program JMP®, Version 15. SAS Institute Inc., Cary, NC, 1989-2020. For detailed information about the countries and the different indexes included in this study, see supplementary file 1.

**RESULTS**

**Hierarchical cluster analysis**

Based on the ability of different countries to control the presentation of new cases and deaths associated with COVID-19 during the first seven months of the pandemic, three main clusters were determined by HCA, and labeled respectively as clusters one, two, and three (Figure 2A).

Cluster one comprised 32.87% of the countries included (Figure 2A) and appeared geographically located at Asia (n=10), North America (n=5), Europe (n=5), Africa (n=4) and Oceania (n=1) (Figure 2A,B). Cluster two contained 48.83% of the total countries included within this study (Figure 2A) and included countries from the Americas (n=11), Asia (n=12), Europe (n=5), and Africa (n=4) (Figure 2A,B). Conversely, countries included in cluster three accounted just for the 21.30% of the countries included in this study (Figure 2A). Geographically, most of these countries were situated in Europe (n=16), and a minority in Asia (n=2), Oceania (n=1) and America (n=1) (Figure 2A,B). All four control countries, Finland, New Zealand, Iceland, and Uruguay, are found within cluster three.

When compared with the other clusters, cluster two showed the average highest positive correlations for all three coefficients (Figure 2A) demonstrating a strong relationship between the variables. This is in contrast with the shape of the graphics produced by countries associated with cluster three, indicating the inability of these countries in cluster two to control the increase in the number of new cases and deaths produced the initial phase of COVID-19. Looking specifically at individual representative countries within the different clusters, the United States (cluster one), Mexico (cluster two), and Italy (cluster three), the differences in numbers of cases and deaths in respect to time can be identified (Figure 2C). Interestingly, when compared with countries from clusters one and three, countries included in cluster number two accounted for the highest total numbers of cases and death after seven months of the pandemic (Figure 2D).

Although cluster one appeared linked to cluster two in terms of hierarchical distance, countries associated with cluster one showed, on average, lower positive correlations for all three coefficients (cases/death, cases/time, death/time). In this sense, the coefficient of correlation associated with the presentation of new deaths through the time was the one with the lowest correlation values, indicating the increased efficiency of these countries to manage with the presentation of new deaths. This situation was consistent when compared the average total numbers of cases and deaths accounted between countries from both clusters (Figure 2D).

Similarly to clusters one and two, average positive correlation values for the coefficient new cases/ new deaths were shown for countries associated for cluster three. However, differing from the other two clusters, low negative correlations were projected between the correlation coefficients new cases/ time and new deaths/ time for countries that integrated in this cluster. As seen in Figure 2C, the shape of the graphic for some of these countries such as Italy, Spain, France and the United Kingdom, demonstrated a dramatic increase of cases and deaths during the first months of the pandemic, a situation that peaked and started decreasing during the subsequent months, indicating the efficacy of these countries to control the contagions and deaths produced by SARS-CoV-2. Furthermore, this situation is reflected in the overall lower number of cases and deaths recorded by countries integrated into cluster three when compared with countries in the other two clusters during the first seven months of the pandemic (Figure 2D).

**Principal Component Analysis**

In general, results by PCA highly correlated with the results as determined by HCA, (Figure 3A). Bartlett test analysis indicated that the two main principal components accounted for

97.167 % of the variability among the countries, being the PC1 responsible for the 71.73% of the total variability (Figure 3B). Spearman's rank test conducted between the coordinates associated with the PC1 and PC2, and the three different correlation coefficients used in this study, indicated that the coefficients new cases/time and new deaths/time were highly correlated with the PC1, while the coefficient new cases/ new deaths was highly correlated with the PC2 (Figure 3C), confirming that the cluster association among different countries depicted in the HCA is highly influenced (71.73%) by the ability of different countries to control the presentations of new cases and deaths through the time.

**Factors influencing the incidence and mortality rates**

To determine the significance of various factors, the Spearman's rank test was used to identify significant ($p<0.0001$) correlations between PC1 (new cases/time and new deaths/time) and the different variables included in this study. Both positive and negative correlations were investigated against all three clusters as defined by HCA. Significant negative correlations were found between PC1 and control of corruption, government effectiveness, HLI, LE, GDP per capita, HDI and the two citation variables included in this study (Figure 4A). These results suggest that the countries demonstrating the highest significance of negative correlation, and which overall were found withing cluster three, showed the lowest correlations in the presentation of new cases and new deaths as based on time. Looking further into specific factors that may have contributed to mortality, the most significant variables that correlated positively with PC1 (Figure 4B) included dietary factors such as deficiencies in Vitamin A, Zinc, and Iron, suggesting the importance of these supplements in the overall outcome of cases. Access to sanitation and availability of handwashing facilities were also positively correlated to number of deaths. These results further support the hierarchical distance between the clusters, namely cluster three in comparison to clusters one and two when compared to the variables presented in this study.

To further assess the dependence among different variables, the Spearman's rank test was conducted between the previously determined indexes which had both positively and negatively correlated with the PC1. Results demonstrated significant correlations ($p<0.0001$), most notably including the HLI, the GDP per capita, and the HDI. Negatively correlated indexes showed the highest correlations against the positively correlated indexes (Figure 4C). Furthermore, comparing positive correlations against each other, significant ($p<0.0001$) correlations were found between variables HLI, HDI and GDP per capita and control of corruption and government effectiveness, demonstrating the dependence among these variables (Figure 4D).

Interestingly, when we conducted this analysis considering just the raw overall total number of cases and deaths produced in each country, no significant correlation was found between these two variables and the multiple indexes considered in this study. Instead, we found that the total number of cases and deaths were positively correlated with differences in live population, gross domestic product, and deaths associated with multiple factors at different countries (figures 5A and B).

**Discussion**

Without doubt, the pandemic produced by SARS-CoV-2 in 2020 is one of the most significant events so far in the history of this century, warranting an investigation that needs to be analyzed from multiple perspectives. Overall, multiple epidemiological studies have focused on explaining the course of this pandemic in terms of the biological properties of SARS-CoV-2 (Bhattacharya et al., 2020; Kumar et al., 2020; Leuzinger et al., 2020), analyses that are highly reasonable considering the unique clinical features manifested by infection with this virus (Sun et al., 2020; Wang et al., 2020). However, based not only on the complex epidemiological triad of SARS-CoV-2 (Shafer and Velazquez-Salinas, 2021), but also in the disparate levels of social inequality worldwide (Thevenot, 2017), we wanted to investigate the existence of potential political, social, and economic factors that might also be driving the course of this pandemic.

In this study, we attempt to retrospectively analyze the first seven months of the pandemic and we consider this first wave of particular interest for several reasons, including the initial shift and centering of this outbreak within affected geographical regions, and subsequent outcomes prior to the distribution of vaccines.

Focusing on an approach allowing us to establish non-parametric correlations between the number of new cases or deaths and the time during the initial outbreak, we were able to evaluate the capabilities of multiple countries to limit both the incidence and mortality produced by SARS-COV-2. Being a novel disease, the lack of general knowledge may have led to a dramatic effect in the increase of the number of cases and deaths at the beginning of the pandemic, as was the case in countries like Italy and Spain. For this reason, the methodology of this study was focused on evaluating the ability of different countries to manage incidence in a timeline manner. While we did take into consideration the evaluation just of the raw number of cases and deaths without including a timeline in their presentation, no significant results were found within the raw data, indicating the independence between both approaches. While others have correlated these variables just with some factors previously described (Clark et al., 2021; Thakur et al., 2021), when time was included as a variable, we were able to correlate these variables with specific economic, political, and social factors. Furthermore, using this approach we were able to correlate incidence and mortality with specific dietary factors deficiencies and sanitary conditions, which were also demonstrated by other studies (D'Amico et al., 2020; Joachimiak, 2021; Li et al., 2020b; Schmidt, 2020), all these issues potentially associated with the susceptibility and persistence of SARS-CoV-2 in the population.

The grouping of countries based on ability to limit incidence and mortality via time, allowed us to directly correlate variables beyond the biological characteristics of SARS-CoV-2. A coordinate system regarding the grouping pattern of different countries using the principal component analysis was determined and these coordinates were used to establish correlations with multiple indexes. In this context, we may consider the extent of the correlations reported herein to be moderate; however, we suggest that the central hypothesis of this study can be supported. Similar studies support our hypothesis involving other viral diseases of public health concern including Ebola (DePinto, 2016; Levi et al., 2018), HIV and influenza(Quinn and Kumar, 2014).

Among the indexes evaluated in this study, we consider that political indexes including correlation of control of corruption and government effectiveness may represent some of the more important influencing variables. In agreement with our findings, numerous situations involving corruption have been described during the pandemic including as the "ignored pandemic" due to its ability to hamper health care systems in light of infectious diseases (Burki, 2019; Garcia, 2019; Teremetskyi et al., 2021; Usman et al., 2022). Corruption has been associated with worldwide immunization progress (Farzanegan and Hofmann, 2021). In fact, corruption has been an issue that has been subject of major concern for the United Nations Office on Drugs and Crime (UNODC) and now it may play a critical role during vaccination around the world (https://www.unodc.org/documents/corruption/COVID-19/Policy_paper_on_COVID-19_vaccines_and_corruption_risks.pdf). In regard to government effectiveness, our results agree with recently published research indicating the impact this may have in the control of mortality (Liang et al., 2020). Indeed, recent studies demonstrate the impact controversial decisions made by governments have had during the course of the pandemic including under-estimating the seriousness of the outbreak, failing in the proper achievement of lockdown measurements, use of facemasks, and undervaluing the importance of diagnostic tests (Agren, 2020; Ibarra-Nava et al., 2020; Ponce, 2020; Taylor, 2020; The Lancet Infectious, 2020).

Furthermore, our analysis showed that there is a high correlation not only between the control of corruption and government effectiveness, but also between these two variables and the other social and economic indexes explored in this study, including HDI (Sarabia et al., 2020), and GDP per capita (Lučić et al., 2016), indicating the importance that corruption and government effectiveness have on human development. Our results, which highlight the correlation of HDI with the control of the pandemic, are consistent with a previous study that indicates that HDI may be an effective marker to assess the progress of the pandemic (Azza and Ayah, 2020).

HDI is a complex concept that is theorized by a series of indicators associated with life expectancy at birth, mean and expected years of schooling, and Gross National Income per capita, explaining the high positive correlations that we found between HDI and the other indexes used in this study. Interestingly, our results also showed the role of GDP per capita as a predictor of the evolution of the pandemic during its initial seven months, being consistent with a previous study in Europe during the first wave of the pandemic (Pardhan and Drydakis, 2020).

Taken together, this study demonstrates an analysis of a variety of significant variables that help define the economic, political, and social factors in influencing the course of incidence rate as well as mortality during the initial phase of the pandemic. However, two important considerations must be made to stress the existence of potential sources of bias in this research, these include the available independent data reported by individual countries as well as the accuracy of early diagnostic results, as well as the emergence of viral variants leading to potential founder effects or natural selection events (Chookajorn, 2020; Daniloski et al., 2021; Grubaugh et al., 2020; Hodcroft et al., 2021; Li et al., 2020a; Velazquez-Salinas et al., 2020; Walensky et al., 2021). Similarly, at this point, we cannot reject the possibility that some correlations reported herein might have been the result of multicollinearity, suggesting that some correlations might not being implying causation, thus results should be interpreted under this precept.

In conclusion, the main purpose of this study was to establish an objective and critical point of view about the multiple factors that may be influencing the control of this pandemic. Herein, we show a general view about the complex situation involving the control of the pandemic of COVID-19 around the world. Our results support the idea that different factors beyond the biological ones may be affecting the control of this pandemic. In this context we suggest that the development of specific mitigation plans to control this pandemic should be made the considering specific political, economic, and social factors prevailing in different geographical regions.

**Credit Author Statement**

**Elizabeth A. Schafer:** Methodology, Investigation, Formal Analysis, Writing- original draft and review & editing. **Samantha Eberl:** Methodology, Data curation, Investigation. **Lauro Velazquez- Salinas:** Conceptualization, Methodology, Investigation, Formal Analysis, Writing- original draft and review & editing, Supervision, Project administration.

**Declaration of competing interest**

The authors have no other declarations of interest.

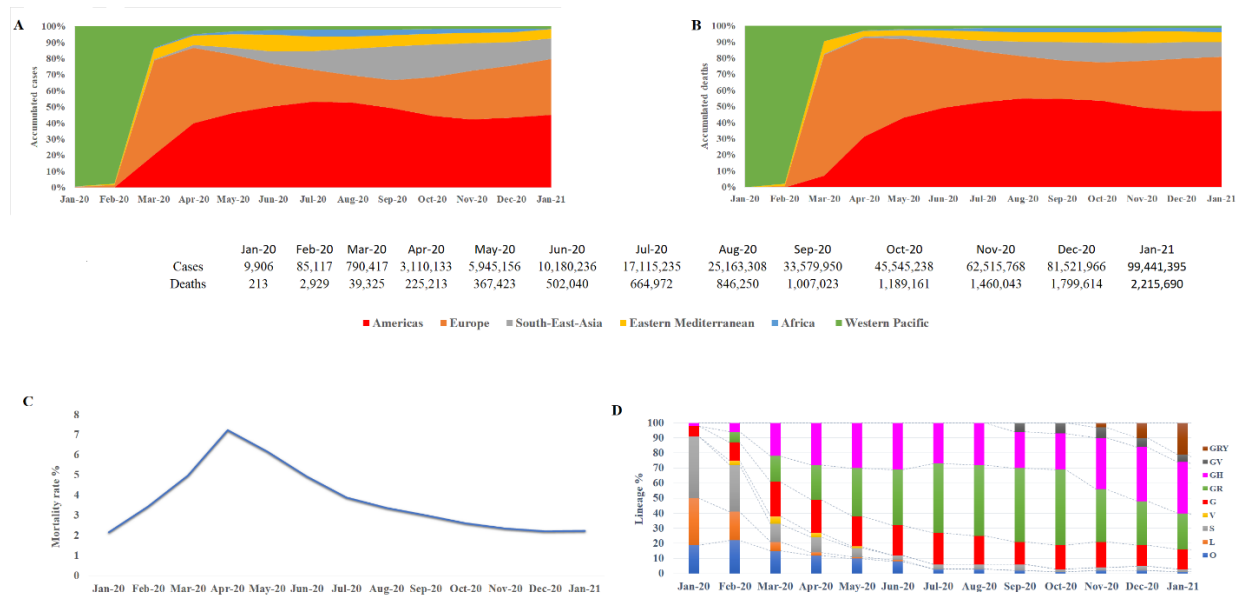

**Figure 1**

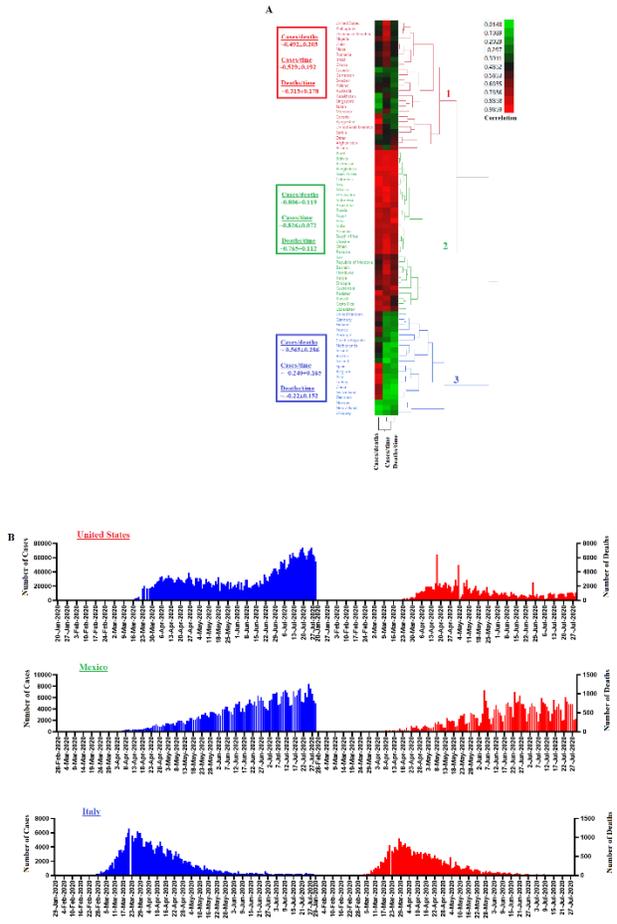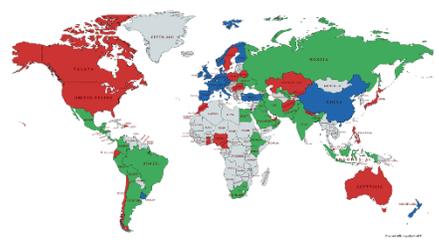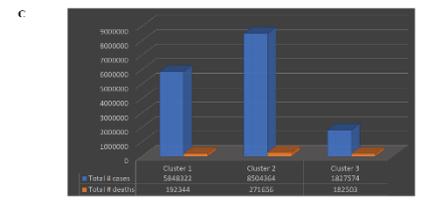

**Figure 2**

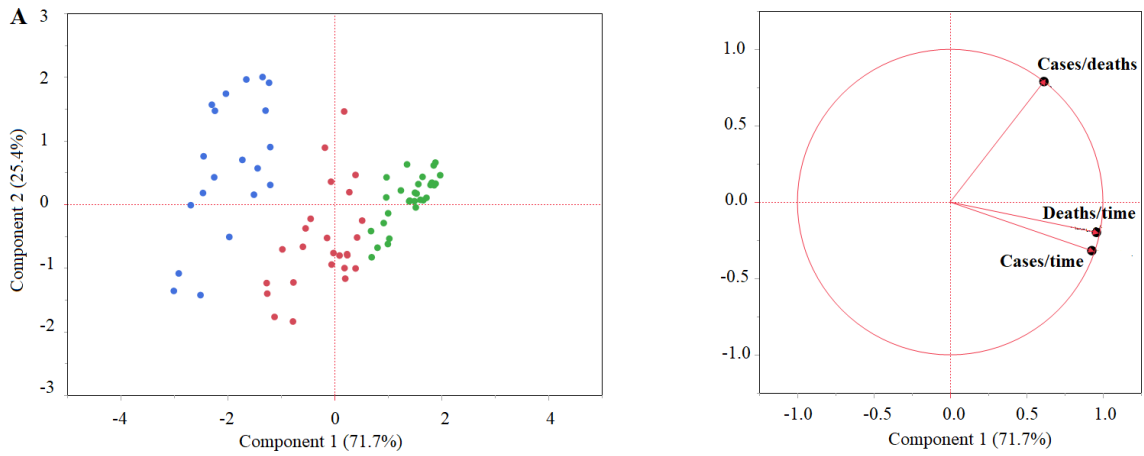

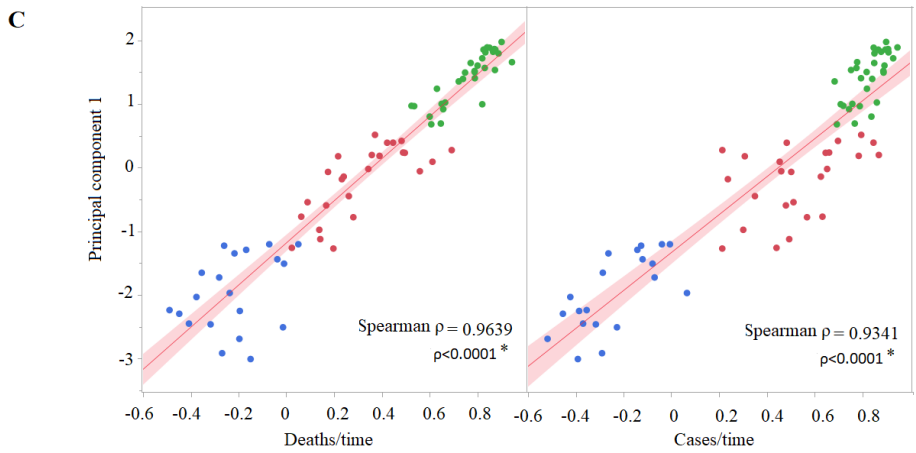

**Figure 3**

A **Principal component 1 correlations**

Negative

| Variable | Spearman ρ | Prob>|ρ| | Cluster 1 | Cluster 2 | Cluster 3 |
|---|---|---|---|---|---|
| Control of corruption | -0.6569 | <.0001 | 0.25 | -0.51 | 1.35 |
| Government effectiveness | -0.6421 | <.0001 | 0.38 | -0.24 | 1.29 |
| HLI | -0.6304 | <.0001 | 64.72 | 60.76 | 76.63 |
| LE | -0.612 | <.0001 | 62.25 | 70.95 | 80.33 |
| GDP per capita ($) | -0.5905 | <.0001 | 22886.99 | 7822.08 | 43802.54 |
| HDI | -0.59 | <.0001 | 0.76 | 0.71 | 0.89 |

B Positive

| Variable | Spearman ρ | Prob>|ρ| | Cluster 1 | Cluster 2 | Cluster 3 |
|---|---|---|---|---|---|
| Zinc deficiency (deaths) | 0.5624 | <.0001 | 268.57 | 285.61 | 8.25 |
| Iron deficiency (deaths) | 0.5577 | <.0001 | 378.61 | 916.04 | 20.97 |
| Vitamin-A deficiency (deaths) | 0.5234 | <.0001 | 1990.05 | 3066.57 | 91.66 |
| Poor sanitation (deaths) | 0.4796 | <.0001 | 3867.70 | 14874.25 | 136.13 |
| No access to handwashing facility (deaths) | 0.4312 | <.0001 | 4643.00 | 12088.05 | 346.27 |

C

| | Zinc deficiency (deaths) | Vitamin-A deficiency (deaths) | Iron deficiency (deaths) | Indoor air pollution (deaths) | Poor sanitation (deaths) | No access to handwashing facility (deaths) |
|---|---|---|---|---|---|---|
| Control of corruption | -0.74 | -0.71 | -0.68 | -0.66 | -0.64 | -0.54 |
| Government effectiveness | -0.72 | -0.69 | -0.68 | -0.64 | -0.63 | -0.53 |
| HLI | -0.83 | -0.82 | -0.80 | -0.74 | -0.76 | -0.67 |
| LE | -0.77 | -0.75 | -0.73 | -0.70 | -0.68 | -0.60 |
| GDP per capita | -0.82 | -0.79 | -0.78 | -0.79 | -0.74 | -0.66 |
| HDI | -0.81 | -0.82 | -0.79 | -0.76 | -0.75 | -0.66 |

D

| | Control of corruption | Government effectiveness | HLI | LE | GDP per capita | HDI |
|---|---|---|---|---|---|---|
| Control of corruption | 1.00 | 0.94 | 0.84 | 0.81 | 0.86 | 0.85 |
| Government effectiveness | 0.94 | 1.00 | 0.86 | 0.83 | 0.89 | 0.89 |
| HLI | 0.84 | 0.86 | 1.00 | 0.97 | 0.91 | 0.94 |
| LE | 0.81 | 0.83 | 0.97 | 1.00 | 0.89 | 0.90 |
| GDP per capita | 0.86 | 0.89 | 0.91 | 0.89 | 1.00 | 0.96 |
| HDI | 0.85 | 0.89 | 0.94 | 0.90 | 0.96 | 1.00 |

# Figure 4

A **Number of cases correlations**

Positive correlations

| Variable | Spearman ρ | Prob>|ρ| |
|---|---|---|
| Population | 0.5993 | <.0001 |
| Outdoor air pollution (deaths) | 0.5782 | <.0001 |
| Drug use (deaths) | 0.5662 | <.0001 |
| Diet low in fiber (deaths) | 0.5488 | <.0001 |
| Diet high in sodium (deaths) | 0.5458 | <.0001 |
| GDP (millions in US $) | 0.5442 | <.0001 |
| High blood pressure (deaths) | 0.5359 | <.0001 |
| Diet low in whole grains (deaths) | 0.534 | <.0001 |
| Diet low in calcium (deaths) | 0.5338 | <.0001 |
| Diet low in nuts and seeds (deaths) | 0.5288 | <.0001 |
| Secondhand smoke (deaths) | 0.5223 | <.0001 |
| Smoking (deaths) | 0.512 | <.0001 |
| Diet low in seafood omega-3 fatty acids (deaths) | 0.5095 | <.0001 |
| Air pollution (outdoor & indoor) (deaths) | 0.5056 | <.0001 |
| Low bone mineral density (deaths) | 0.4862 | <.0001 |
| Diet low in legumes (deaths) | 0.4477 | <.0001 |

B **Number of deaths correlations**

Positive correlations

| Variable | Spearman ρ | Prob>|ρ| |
|---|---|---|
| Diet low in calcium (deaths) | 0.6562 | <.0001 |
| Diet low in fiber (deaths) | 0.642 | <.0001 |
| Low bone mineral density (deaths) | 0.6349 | <.0001 |
| Population | 0.631 | <.0001 |
| Smoking (deaths) | 0.6272 | <.0001 |
| Drug use (deaths) | 0.6213 | <.0001 |
| GDP (millions in US $) | 0.6208 | <.0001 |
| Outdoor air pollution (deaths) | 0.6191 | <.0001 |
| High blood pressure (deaths) | 0.6091 | <.0001 |
| Diet high in sodium (deaths) | 0.5965 | <.0001 |
| Diet low in nuts and seeds (deaths) | 0.5855 | <.0001 |
| Diet low in whole grains (deaths) | 0.5835 | <.0001 |
| Secondhand smoke (deaths) | 0.5762 | <.0001 |
| Diet low in seafood omega-3 fatty acids (deaths) | 0.5512 | <.0001 |
| Air pollution (outdoor & indoor) (deaths) | 0.55 | <.0001 |
| Diet low in legumes (deaths) | 0.5246 | <.0001 |
| Alcohol use (deaths) | 0.5025 | <.0001 |

Figure 5

Legends

**Figure 1. Overview of International Geographic Data within Initial Seven-Month Phase of Pandemic, 2020.** Graphics show the percentage of accumulated cases **(A)** and deaths **(B)** among different geographic regions, represented by different colors. The total number of cases and deaths is displayed below the graphics. Information was obtained from the world Health Organization database. **C)** Shows the overall monthly mortality rate around the world produced by SARS-CoV-2 during 2020. **D)** Proportion of genetic lineages (GISAID classification) of SARS-CoV-2 recovered monthly from patients infected of COVID-19 around the world during 2020. Information was obtained from the GISAID database.

**Figure 2. Ability of different countries to deal with the presentation of new cases and deaths during the first waive of the pandemic.** **A)** Hierarchical cluster analysis shows the grouping pattern of different countries included in this study based on the Spearman correlation obtained for the variables cases/deaths, cases/time, and deaths/time. The spectrum of colors between green and read show the intensity of the correlation among different variables. The average correlation of these variables for each cluster is shown in the squares. B) Geographical distribution of associated with clusters one (red), two (green) and three (blue) as determined by HCA. **C)** Pattern of distribution of new cases and new deaths of United States, Mexico, and Italy, representing countries associated with clusters one, two, and three, respectively. **D)** Comparison between the total number of cases and deaths among countries associated with different clusters.

**Figure 3. Effect of different variables in the grouping pattern among countries. A)** Principal component analysis was conducted to assess the influence of variables cases/deaths, cases/time, and deaths/time in the grouping pattern of different countries. Red, green, and blue dots represent countries associated with previously determined clusters one, two and three respectively**. B)** Bartlett test analysis reflecting the reliability of principal components one and two. **C)** Spearman correlation analysis showing the high correlation of variables cases/time, and deaths time with the principal component 1.

**Figure 4. Correlation of the principal component 1 with multiple social, economic, and politic indexes and different causes of deaths.** Principal component one coordinates were used to establish Spearman correlations with different indexes and causes of deaths. The analysis found statistically significant negative **(A)** and positive correlations **(B)**. Average values of indexes and numbers of different causes of deaths for different clusters comprising different countries are shown. The dependence between factors found negatively and positively correlated **(C),** as well within factors negatively **(D)** correlated with the first principal component were calculated by Spearman correlation analyses. In all cases red and blue colors represent negative and positive correlations respectively. *NS = Not Significant

**Figure 5. Correlations between total number of cases and deaths with multiple factors.** The raw total number of cases **(A)** and deaths **(B)** were used to establish Spearman correlations among different variables and causes of deaths.

**Supplementary file 1.** Data information.

**References**

Agren, D., 2020. Mexican President Lopez Obrador draws doctors' ire. Lancet 395, 1601.

Alanezi, F., Aljahdali, A., Alyousef, S.M., Alrashed, H., Mushcab, H., AlThani, B., Alghamedy, F., Alotaibi, H., Saadah, A., Alanzi, T., 2020. A Comparative Study on the Strategies Adopted by the United Kingdom, India, China, Italy, and Saudi Arabia to Contain the Spread of the COVID-19 Pandemic. J Healthc Leadersh 12, 117-131.

Alfano, V., Ercolano, S., 2020. The Efficacy of Lockdown Against COVID-19: A Cross-Country Panel Analysis. Appl Health Econ Health Policy 18, 509-517.

Armocida, B., Formenti, B., Ussai, S., Palestra, F., Missoni, E., 2020. The Italian health system and the COVID-19 challenge. Lancet Public Health 5, e253.

Azza, A.-F., Ayah, S., 2020. Using Human Development Indices to Identify Indicators to Monitor the Corona Virus Pandemic. Journal of Current Viruses and Treatment Methodologies 1, 48-57.

Bartlett, M.S., 1937. Properties of Sufficiency and Statistical Tests. Proceedings of the Royal Society of London. Series A, Mathematical and Physical Sciences 160, 268-282.

Bhattacharya, S., Basu, P., Poddar, S., 2020. Changing epidemiology of SARS-CoV in the context of COVID-19 pandemic. J Prev Med Hyg 61, E130-E136.

Bishara, A.J., Hittner, J.B., 2017. Confidence intervals for correlations when data are not normal. Behavior Research Methods 49, 294-309.

Burki, T., 2019. Corruption is an "ignored pandemic". Lancet Infect Dis 19, 471.
Chookajorn, T., 2020. Evolving COVID-19 conundrum and its impact. Proc Natl Acad Sci U S A 117, 12520-12521.

Clark, C.E., McDonagh, S.T.J., McManus, R.J., Martin, U., 2021. COVID-19 and hypertension: risks and management. A scientific statement on behalf of the British and Irish Hypertension Society. J Hum Hypertens.

Cousins, S., 2020. New Zealand eliminates COVID-19. Lancet 395, 1474.

D'Amico, F., Peyrin-Biroulet, L., Danese, S., 2020. Oral Iron for IBD Patients: Lessons Learned at Time of COVID-19 Pandemic. J Clin Med 9.

Daniloski, Z., Jordan, T.X., Ilmain, J.K., Guo, X., Bhabha, G., tenOever, B.R., Sanjana, N.E., 2021. The Spike D614G mutation increases SARS-CoV-2 infection of multiple human cell types. Elife 10.

DePinto, J.R., 2016. Corruption and the 2014 EVD Crisis in Sierra Leone: Ebola as "Total Disease", in: Mustapha, M., Bangura, J.J. (Eds.), Democratization and Human Security in Postwar Sierra Leone. Palgrave Macmillan US, New York, pp. 217-249.

Farzanegan, M.R., Hofmann, H.P., 2021. Effect of public corruption on the COVID-19 immunization progress. Sci Rep 11, 23423.

Garcia, P.J., 2019. Corruption in global health: the open secret. Lancet 394, 2119-2124.
Ghislandi, S., Sanderson, W.C., Scherbov, S., 2019. A Simple Measure of Human Development: The Human Life Indicator. Popul Dev Rev 45, 219-233.

Grubaugh, N.D., Petrone, M.E., Holmes, E.C., 2020. We shouldn't worry when a virus mutates during disease outbreaks. Nat Microbiol 5, 529-530.

Gudbjartsson, D.F., Helgason, A., Jonsson, H., Magnusson, O.T., Melsted, P., Norddahl, G.L., Saemundsdottir, J., Sigurdsson, A., Sulem, P., Agustsdottir, A.B., Eiriksdottir, B., Fridriksdottir, R., Gardarsdottir, E.E., Georgsson, G., Gretarsdottir, O.S., Gudmundsson, K.R., Gunnarsdottir, T.R., Gylfason, A., Holm, H., Jensson, B.O., Jonasdottir, A., Jonsson, F., Josefsdottir, K.S., Kristjansson, T., Magnusdottir, D.N., le Roux, L., Sigmundsdottir, G., Sveinbjornsson, G., Sveinsdottir, K.E., Sveinsdottir, M., Thorarensen, E.A., Thorbjornsson, B., Love, A., Masson, G., Jonsdottir, I., Moller, A.D., Gudnason, T., Kristinsson, K.G., Thorsteinsdottir, U., Stefansson, K., 2020. Spread of SARS-CoV-2 in the Icelandic Population. N Engl J Med 382, 2302-2315.

Hodcroft, E.B., Zuber, M., Nadeau, S., Vaughan, T.G., Crawford, K.H.D., Althaus, C.L., Reichmuth, M.L., Bowen, J.E., Walls, A.C., Corti, D., Bloom, J.D., Veesler, D., Mateo, D., Hernando, A., Comas, I., Gonzalez-Candelas, F., Seq, C.-S.c., Stadler, T., Neher, R.A., 2021. Spread of a SARS-CoV-2 variant through Europe in the summer of 2020. Nature 595, 707-712.

Ibarra-Nava, I., Cardenas-de la Garza, J.A., Ruiz-Lozano, R.E., Salazar-Montalvo, R.G., 2020. Mexico and the COVID-19 Response. Disaster Med Public Health Prep 14, e17-e18.

Joachimiak, M.P., 2021. Zinc against COVID-19? Symptom surveillance and deficiency risk groups. PLoS Negl Trop Dis 15, e0008895.

Kumar, M., Taki, K., Gahlot, R., Sharma, A., Dhangar, K., 2020. A chronicle of SARS-CoV-2: Part-I - Epidemiology, diagnosis, prognosis, transmission and treatment. Sci Total Environ 734, 139278.

Leuzinger, K., Roloff, T., Gosert, R., Sogaard, K., Naegele, K., Rentsch, K., Bingisser, R., Nickel, C.H., Pargger, H., Bassetti, S., Bielicki, J., Khanna, N., Tschudin Sutter, S., Widmer, A., Hinic, V., Battegay, M., Egli, A., Hirsch, H.H., 2020. Epidemiology of Severe Acute Respiratory

Syndrome Coronavirus 2 Emergence Amidst Community-Acquired Respiratory Viruses. J Infect Dis 222, 1270-1279.

Levi, J., Pozniak, A., Heath, K., Hill, A., 2018. The impact of HIV prevalence, conflict, corruption, and GDP/capita on treatment cascades: data from 137 countries. J Virus Erad 4, 80-90.

Li, Q., Wu, J., Nie, J., Zhang, L., Hao, H., Liu, S., Zhao, C., Zhang, Q., Liu, H., Nie, L., Qin, H., Wang, M., Lu, Q., Li, X., Sun, Q., Liu, J., Zhang, L., Li, X., Huang, W., Wang, Y., 2020a. The Impact of Mutations in SARS-CoV-2 Spike on Viral Infectivity and Antigenicity. Cell 182, 1284-1294 e1289.

Li, R., Wu, K., Li, Y., Liang, X., Tse, W.K.F., Yang, L., Lai, K.P., 2020b. Revealing the targets and mechanisms of vitamin A in the treatment of COVID-19. Aging (Albany NY) 12, 15784-15796.

Liang, L.L., Tseng, C.H., Ho, H.J., Wu, C.Y., 2020. Covid-19 mortality is negatively associated with test number and government effectiveness. Sci Rep 10, 12567.

Lučić, D., Radišić, M., Dobromirov, D., 2016. Causality between corruption and the level of GDP. Economic Research-Ekonomska Istraživanja 29, 360-379.

Mazzucchelli, R., Agudo Dieguez, A., Dieguez Costa, E.M., Crespi Villarias, N., 2020. [Democracy and Covid-19 mortality in Europe.]. Rev Esp Salud Publica 94.

Moreno, P., Moratorio, G.A., Iraola, G., Fajardo, A., Aldunate, F., Pereira, M., Perbolianachis, P., Costabile, A., Lopez-Tort, F., Simon, D., Salazar, C., Ferres, I., Diaz-Viraque, F., Abin, A., Bresque, M., Fabregat, M., Maidana, M., Rivera, B., Cruces, M., Rodriguez, J., Scavone, P., Alegretti, M., Nabon, A., Gagliano, G., Rosa, R., Henderson, E., Bidegain, E., Zarantonelli, L., Piattoni, C., Greif, G., Francia, M., Robello, C., Duran, R., Brito, G., Bonnecarrere, V., Sierra, M., Colina, R., Marin, M., Cristina, J., Erlich, R., Paganini, F., Cohen, H., Radi, R., Barbeito, L., Badano, J., Pritsch, O., Fernandez, C., Arim, R., Batthyany, C., 2020. An effective COVID-19 response in South America: the Uruguayan Conundrum. medRxiv, 2020.2007.2024.20161802.

Ojong, N., 2020. The COVID-19 Pandemic and the Pathology of the Economic and Political Architecture in Cameroon. Healthcare (Basel) 8.

Pardhan, S., Drydakis, N., 2020. Associating the Change in New COVID-19 Cases to GDP per Capita in 38 European Countries in the First Wave of the Pandemic. Front Public Health 8, 582140. Penny, K.I., Jolliffe, I.T., 1999. Multivariate outlier detection applied to multiply imputed laboratory data. Stat Med 18, 1879-1895; dicussion 1897.

Ponce, D., 2020. The impact of coronavirus in Brazil: politics and the pandemic. Nat Rev Nephrol 16, 483.

Quinn, S.C., Kumar, S., 2014. Health inequalities and infectious disease epidemics: a challenge for global health security. Biosecur Bioterror 12, 263-273.

Sarabia, M., Crecente, F., del Val, M.T., Giménez, M., 2020. The Human Development Index (HDI) and the Corruption Perception Index (CPI) 2013-2017: analysis of social conflict and populism in Europe. Economic Research-Ekonomska Istraživanja 33, 2943-2955.

Schmidt, C.W., 2020. Lack of Handwashing Access: A Widespread Deficiency in the Age of COVID-19. Environ Health Perspect 128, 64002.

Shafer, A.E., Velazquez-Salinas, L., 2021. Controlling the COVID-19 Pandemic: The complex Epidemiological Triad of SARS-CoV-2. Int. Jr. Infect Dis &Epidemlgy 2, 41-42.

Sun, P., Qie, S., Liu, Z., Ren, J., Li, K., Xi, J., 2020. Clinical characteristics of hospitalized patients with SARS-CoV-2 infection: A single arm meta-analysis. J Med Virol 92, 612-617.

Taylor, L., 2020. Covid-19: How denialism led Mexico's disastrous pandemic control effort. BMJ 371, m4952.

Teremetskyi, V., Duliba, Y., Kroitor, V., Korchak, N., Makarenko, O., 2021. Corruption and strengthening anti-corruption efforts in healthcare during the pandemic of Covid-19. Med Leg J 89, 25-28.

Thakur, B., Dubey, P., Benitez, J., Torres, J.P., Reddy, S., Shokar, N., Aung, K., Mukherjee, D., Dwivedi, A.K., 2021. A systematic review and meta-analysis of geographic differences in comorbidities and associated severity and mortality among individuals with COVID-19. Sci Rep 11, 8562.

The Lancet Infectious, D., 2020. Political casualties of the COVID-19 pandemic. Lancet Infect Dis 20, 755.

Thevenot, C., 2017. Inequality in OECD countries. Scand J Public Health 45, 9-16.

Usman, M., Husnain, M., Akhtar, M.W., Ali, Y., Riaz, A., Riaz, A., 2022. From the COVID-19 pandemic to corrupt practices: a tale of two evils. Environ Sci Pollut Res Int.

Velazquez-Salinas, L., Zarate, S., Eberl, S., Gladue, D.P., Novella, I., Borca, M.V., 2020. Positive Selection of ORF1ab, ORF3a, and ORF8 Genes Drives the Early Evolutionary Trends of SARS-CoV-2 During the 2020 COVID-19 Pandemic. Frontiers in Microbiology 11.

Velazquez-Salinas, L., Zarate, S., Eschbaumer, M., Pereira Lobo, F., Gladue, D.P., Arzt, J., Novella, I.S., Rodriguez, L.L., 2016. Selective Factors Associated with the Evolution of Codon Usage in Natural Populations of Arboviruses. PLoS One 11, e0159943.

Walensky, R.P., Walke, H.T., Fauci, A.S., 2021. SARS-CoV-2 Variants of Concern in the United States-Challenges and Opportunities. JAMA.

Wang, Y., Wang, Y., Chen, Y., Qin, Q., 2020. Unique epidemiological and clinical features of the emerging 2019 novel coronavirus pneumonia (COVID-19) implicate special control measures. J Med Virol 92, 568-576.

Wu, F., Zhao, S., Yu, B., Chen, Y.M., Wang, W., Song, Z.G., Hu, Y., Tao, Z.W., Tian, J.H., Pei, Y.Y., Yuan, M.L., Zhang, Y.L., Dai, F.H., Liu, Y., Wang, Q.M., Zheng, J.J., Xu, L., Holmes, E.C., Zhang, Y.Z., 2020. A new coronavirus associated with human respiratory disease in China. Nature 579, 265-269.

Ye, Z.W., Yuan, S., Yuen, K.S., Fung, S.Y., Chan, C.P., Jin, D.Y., 2020. Zoonotic origins of human coronaviruses. Int J Biol Sci 16, 1686-1697.